\documentclass[aps,prc,showpacs,twocolumn,letter]{revtex4}

\usepackage{graphicx,epsfig}

\begin{document}

\title{On the Possibility of Event Shape Selection in Relativistic Heavy Ion Collisions}

\author{Hannah Petersen}
\affiliation{Frankfurt Institute for Advanced Studies, Ruth-Moufang-Str. 1, D-60438 Frankfurt am Main, Germany}
\author{Berndt M\"uller}
\affiliation{Department of Physics, Duke University, Durham, North Carolina 27708-0305, USA}


\begin{abstract}
We investigate the possibility of selecting heavy ion collision events with certain features in the initial state (``event engineering''). Anisotropic flow measurements in heavy ion reactions have confirmed the almost ideal fluid dynamical behaviour of the hot and dense quark gluon plasma state. As a consequence, it is intriguing to pursue the idea of selecting collisions with a certain special initial geometry, e.g., a large ellipsoidal or triangular deformation, by classifying events by the value of their final observed flow coefficients. This procedure could be especially interesting for azimuthally dependent jet energy loss studies. We investigate the correlation between initial state features and final state momentum space anisotropies within an event-by-event hybrid approach. We find that the finite particle number and hadronic rescattering of the final state leads to large event-by-event fluctuations in the observables that could be used to characterize the features of the initial state. This makes event engineering by final state selection difficult.
\end{abstract}



\maketitle

\section{Introduction}
In recent years, fluctuations in the geometry of the initial state of a relativistic heavy ion collision have attracted much interest \cite{Alver:2010gr,Petersen:2010cw,Schenke:2010rr,Adare:2012kf}. The experimental approach involves the investigation of higher moments of the azimuthal momentum distribution of the particles produced in the collision \cite{Adare:2011tg,ALICE:2011ab,Adamczyk:2013waa}. Especially for the two largest coefficients in the Fourier expansion of the azimuthal particle distribution, $v_2$ and $v_3$, it has been demonstrated that there exists a rather linear correlation with the initial state eccentricity $\epsilon_2$ and triangularity $\epsilon_3$ \cite{Teaney:2010vd,Gardim:2011xv}. 

The next logical step would be to measure the event-by-event distributions of the flow coefficients \cite{Aad:2013xma}. These distributions might be useful to disentangle the effect of a finite shear viscosity over entropy ratio and the size of fluctuations in the initial energy density profiles. The mean value of the flow coefficients seems to be determined by the hydrodynamic response while the shape is related to the initial eccentricity distribution from a given initial state parametrization \cite{Niemi:2012aj,Gale:2012rq}. 

Extending the exploration of final state correlations in terms of anisotropic flow measurements to initial state features, the idea of ``event shape engineering'' emerged \cite{Schukraft:2012ah}. Usually events are classified by their centrality to select rather head-on versus more peripheral collisions. To have a better handle on observables that depend on the details of the geometry of the initial state, like event plane dependent energy loss or azimuthally dependent HBT radii, it is intriguing to select events by their $v_2$ or $v_3$ values to constrain the initial geometry of the sample. 

In this paper the selection of initial event shapes by final state measurable flow coefficients is explored in a three-dimensional event-by-event hybrid approach. The present study is intended to provide a qualitative estimate of the possibility of event shape engineering in single events. In Section \ref{model} the hybrid approach and the setup for this analysis is described. The correlation of centrality selection and final state flow coefficients is studied in Section \ref{centrality}. Section \ref{inigeo} contains the main results on geometry selection by event shape engineering. The conclusions are presented in the last Section \ref{concl}.

\section{Model Description}
\label{model}

To obtain a realistic estimate on how well one is able to constrain the initial state geometry by selecting certain values of final state anisotropic flow values an event-by-event hybrid approach must be employed. Hybrid approaches based on nearly ideal fluid dynamics for the hot and dense stage and hadron transport for the later dilute stage of the reaction have been applied successfully to describe the bulk properties of heavy ion collisions at RHIC and LHC. 

For our analysis we used the hybrid approach UrQMD-3.1p1 \cite{Bass:1998ca,Bleicher:1999xi,Petersen:2008dd} to simulate Au+Au collisions at $\sqrt{s_{\rm NN}}=200$ GeV for a range of impact parameters $b=0-12$ fm. In this model the initial state is generated by the UrQMD transport approach. At the highest RHIC energy most of the initial binary nucleon-nucleon collisions result in hard interactions that are treated with PYTHIA \cite{Sjostrand:2006za}. At an initial start time of $t_{\rm start}=0.5$ fm, all the particles that have participated in at least one interaction and are at rapidities $|y|<2$ are converted to energy, momentum and net baryon number densities with three-dimensional gaussian distributions (width $\sigma=1$ fm) that are Lorentz contracted along the beam direction. 

The ideal hydrodynamic evolution is solved using the SHASTA algorithm \cite{Rischke:1995ir,Rischke:1995mt} with an equation of state that is based on a chiral Lagrangian coupled to the Polyakov loop (DE-EoS) and fits the lattice data at zero baryochemical potential \cite{Steinheimer:2011ea}. To switch back to particle degrees of freedom an approximate iso-$\tau$ hypersurface is constructed by converting transverse slices once all cells within a slice have cooled below the energy density criterion of $7 \epsilon_0$ where $\epsilon_0$ is the nuclear ground state energy density \cite{Petersen:2009mz}. After this transition the hadronic rescattering and decays are treated with the hadron transport approach UrQMD.

This model with exactly the same parameters as quoted above has been shown to reproduce particle yields and spectra at RHIC \cite{Petersen:2010zt}. Since we have the information about the full evolution available, this approach is well suited to study the correlation between initial state properties and final state observables. In addition, the final state consists of a finite number of particles for each event, so that the same analysis procedures as used in experiments can be applied. Currently, this approach is limited to ideal hydrodynamics for the hot and dense stage of the evolution. How a finite viscosity for the quark gluon plasma would influence the correlations is left for future investigations.

\section{Centrality Selection}
\label{centrality}

\begin{figure}[ht]
\resizebox{0.5\textwidth}{!}{ 
\centering
\includegraphics{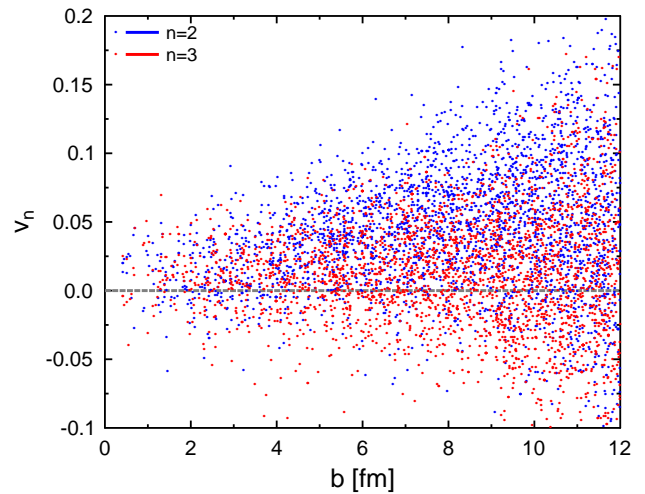}
}
\caption{(Color online) Scatter plot of $v_2$ and $v_3$ as a function of impact parameter $b$.} 
\label{fig_vn_imp}
\end{figure}

\begin{figure}[ht]
\resizebox{0.5\textwidth}{!}{ 
\centering
\includegraphics{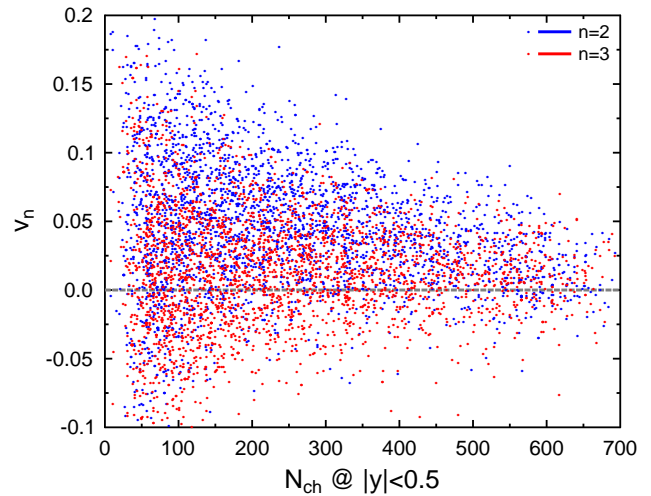}
}
\caption{(Color online) Scatter plot of $v_2$ (blue dots) and $v_3$ (red dots) as a function of charged particle yield at midrapidity.} 
\label{fig_vn_centrality}
\end{figure}

The centrality of a heavy ion collision is quantified by the impact parameter in theoretical calculations or by the charged particle multiplicity in experimental investigations. In Figs. \ref{fig_vn_imp} and \ref{fig_vn_centrality} the final state elliptic and triangular flow values, $v_2$ and $v_3$, are shown as a function of these two quantities. Each point on the scatter plot represents one event. The anisotropic flow values have been calculated using the event plane method
\begin{equation}
v_n=\langle \cos n (\phi-\Psi_n)\rangle \quad \mbox{with } \Psi_n=\tan^{-1} \frac{\langle p_T \sin(n\phi)\rangle}{\langle p_T \cos(n \phi) \rangle} 
\end{equation}
where $\langle \dot \rangle$ denotes the average over all particles at midrapidity in one event and $\phi$ is the azimuthal angle in momentum space. The resolution correction has not been applied, since our calculation is based on the final state particle distributions from single events. Auto-correlations are avoided in the $v_n$ calculation by ensuring that the particle that is correlated with the event plane has not been considered in the corresponding determination of the plane itself. In Fig. \ref{fig_v2dist} this method has been compared to the elliptic flow calculated with the known reaction plane and the results are very similar. As in the literature, the event plane definition contains a transverse momentum weight, which is not present in the definition of $v_n$, therefore negative values for the flow coefficients can occur.  

The fluctuations of the flow coefficients are larger for more peripheral events as expected, since there are fewer particles produced. For elliptic flow $v_2$ the characteristic centrality dependence of the mean value that is small in central collisions and visibly increases towards mid-central collisions can be seen. However, the scatter plots show clearly that at any impact parameter one observes a wide range of values for the anisotropic flow coefficients, Therefore, a precise geometry selection is impossible by dividing events into centrality classes.  

\begin{figure}[ht]
\resizebox{0.4\textwidth}{!}{ 
\centering
\includegraphics{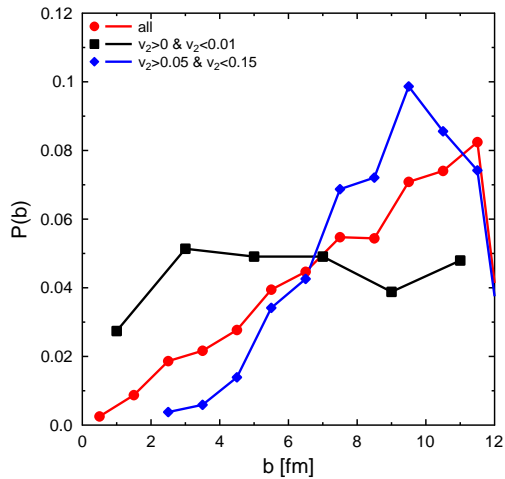}
}
\caption{(Color online) Distribution of impact parameters under certain elliptic flow constraints.} 
\label{fig_distb}
\end{figure}

In Fig.~\ref{fig_distb} this finding is presented in a different way. As a reference the usual linear distribution of events as a function of impact parameter is displayed by the red circles and line. The blue diamonds depict the impact parameter distribution of events with large elliptic flow values $0.15>v_2>0.05$, which is seen to be tilted toward higher impact parameters. The selection of small elliptic flow values $0<v_2<0.01$ results in a much flatted impact parameter distribution as shown by the black squares. We conclude that selecting events with certain $v_2$ ranges leads to broad or even flat distributions as a function of centrality, and one can find events with specific flow anisotropies in all centrality classes. 

\section{Constraints on Initial Geometry}
\label{inigeo}

\begin{figure}[t]
\resizebox{0.4\textwidth}{!}{ 
\centering
\includegraphics{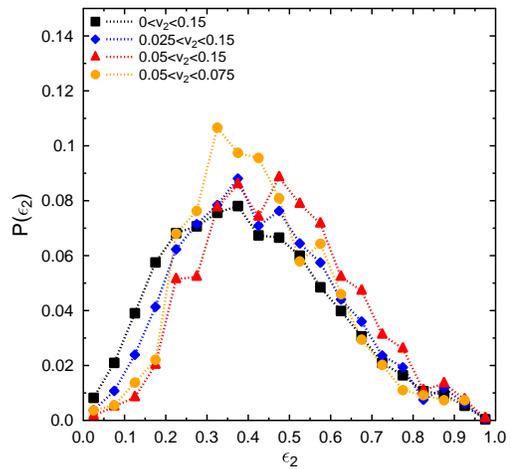}
}
\caption{(Color online) Probability distributions of the initial eccentricity in events with high elliptic flow values. } 
\label{fig_probepshigh}
\end{figure}

We now explore how much the initial event geometry can be constrained by choosing only events in a specific range of elliptic flow values $v_2$. We have also studied event selection by constraining triangular flow values $v_3$, but could not find significant effects on the initial triangularity distributions. Fig. \ref{fig_probepshigh} shows the probability distribution of eccentricities $\epsilon_2$ in minimum bias events with certain cuts on the final elliptic flow values. The black squares represent the distribution that is very close to the original distribution for all events ($0<v_2<0.15$). One observes that the exclusion of low $v_2$ values reduces the probability for small eccentricities considerably, while cutting out the higher $v_2$ events does not reduce the range of initial geometries much.  

\begin{figure}[t]
\resizebox{0.4\textwidth}{!}{ 
\centering
\includegraphics{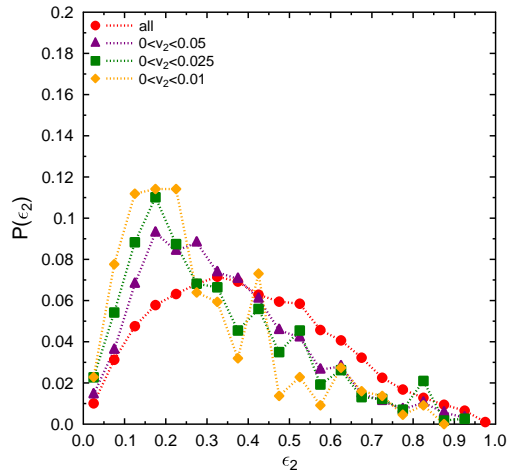}
}
\caption{(Color online) Probability distributions of the initial 
eccentricity in events with low (right) elliptic flow 
values. } 
\label{fig_probepssmall}
\end{figure}

Selecting events where elliptic flow is smaller modifies the shape of the eccentricity distribution towards lower values as expected, as shown in Fig. \ref{fig_probepssmall}, but even for very small $v_2$ (yellow diamonds) the peak of the eccentricity distribution is at nonzero $\epsilon_2$. Overall, one can see that event selection by restrictions on elliptic flow changes the distribution of initial state eccentricities, but they remain rather broad and not narrow enough to select a well defined initial state configuration.  

Since fluctuations in the initial state energy density distributions may be associated with the production of particles at high transverse momentum $p_t$, we have checked that requiring one particle above a certain transverse momentum cut of $p_t > 5,10,20$ GeV/c does not change the conclusions. 

\begin{figure}[ht]
\resizebox{0.4\textwidth}{!}{ 
\centering
\includegraphics{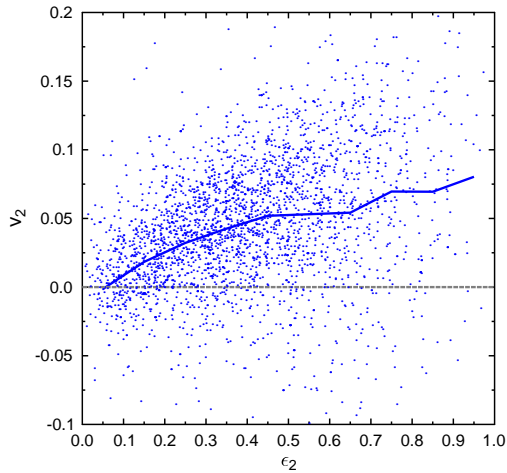}
}
\caption{(Color online) Scatter plot of $v_2$($\epsilon_2$) within the 
hybrid approach, the mean is indicated by the full line.} 
\label{fig_v2scatter}
\end{figure}

\begin{figure}[ht]
\resizebox{0.4\textwidth}{!}{ 
\centering
\includegraphics{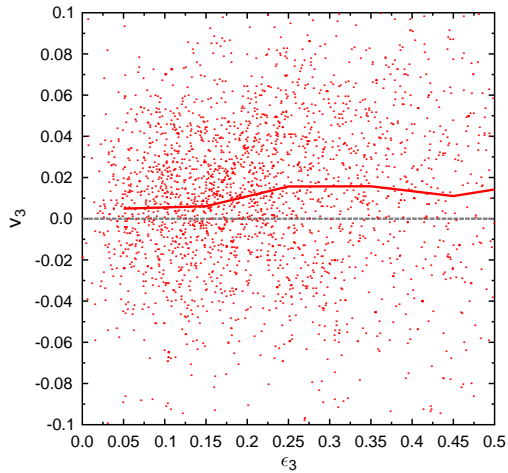}
}
\caption{(Color online) Scatter plots of $v_3$($\epsilon_3$) within the 
hybrid approach, the mean value is depicted by the full line. } 
\label{fig_v3scatter}
\end{figure}

The scatter plots shown in Figs. \ref{fig_v2scatter} and \ref{fig_v3scatter} are helpful to examine the correlation between final state flow observables and initial state eccentricities. The full lines showing the mean values exhibit the expected approximately linear relationship between $v_2$ ($v_3$) and $\epsilon_2$ ($\epsilon_3$). One the other hand, the fluctuations imply very large ranges of initial eccentricity or triangularity that correspond to selected elliptic or triangular flow values. These scatter plots correspond to minimum bias events and at least for elliptic flow the correlation might be stronger in mid-central collisions. But as we have shown in Figs. \ref{fig_vn_imp} and \ref{fig_vn_centrality} the selection of a specific centrality class does not reduce much the range of observed flow coefficients.

A strict linear correlation of initial state eccentricity to final state flow anisotropy would be expected from ideal hydrodynamics. Then what causes the wide range of $v_2 (v_3)$ values corresponding to a fixed $\epsilon_2 (\epsilon_3)$? The major source of these broad distributions in our model is the sampling of final state particles and rescattering that is included in the hybrid approach. On one hand, this treatment of the freeze-out process is more realistic than assuming free-streaming particles directly after the ideal hydrodynamical evolution. On the other hand, in our case, a finite number of particles is sampled only once per hydrodynamic evolution in contrast to the oversampling that has been applied in other hybrid approaches \cite{Qiu:2011iv}. In reality, the momentum correlations that are built up during the evolution correlated to initial spatial correlations will be carried through the whole evolution of the fireball. Since in our model hydrodynamics is applied to the intermediate phase of the reaction, the microscopic information present in this phase is coarse-grained and cannot be faithfully recovered by a single sample of hadrons on the hydro-to-micro conversion hypersurface. This is an inherent limitation of the hybrid approach.

\begin{figure}[ht]
\resizebox{0.4\textwidth}{!}{ \centering
\includegraphics{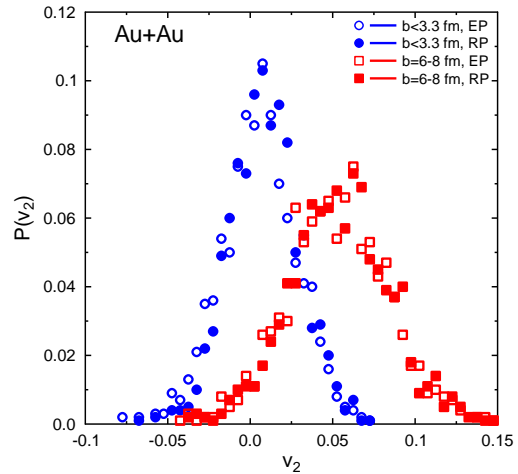}
}
\caption{(Color online) $v_2$ distribution for one smooth event after sampling and propagation in hadron cascade calculated with two different methods.} 
\label{fig_v2dist}
\end{figure}

To demonstrate that the final state particle sampling and rescattering is a major source of broadening the $v_2$ distribution, we have calculated the distribution of final state elliptic flow values for a test case. Smooth initial conditions have been constructed by averaging over 100 initial states for central ($b<3.3$ fm) and mid-central ($b=6-8$ fm) collisions. The ideal hydrodynamical evolution was performed only once, but then many final states were generated by multiple samplings and final state rescattering calculations. Within the framework of hydrodynamics alone, one would expect a well-defined elliptic flow result, but due to the finite particle multiplicity the hybrid approach yields a normal distribution around the expected value as shown in Fig. \ref{fig_v2dist}. In order to demonstrate that the resolution correction, which cannot be applied event by event, has no influence on our result, two different methods of calculating $v_2$ are compared. The open symbols represent the results applying the event plane method, while the full symbols correspond to the results using the reaction plane method where the elliptic flow coefficient is defined as $v_2=\langle (p_x^2-p_y^2)/(p_x^2+p_y^2)\rangle$. As one sees, the two methods yield the same distributions.

\section{Conclusion}
\label{concl}

In summary, we explored the idea of constraining initial event shapes by selecting final state anisotropic flow coefficients in individual collision events. Our investigation was carried out in the framework of an event-by-event hybrid approach for the heavy ion reaction. We found that the final state of the quark-gluon fluid into a finite number of hadrons and their propagation in a Boltzmann transport model results in very broad distributions of elliptic and triangular flow values for any given value of the initial geometrical eccentricity and triangularity. It would be interesting to explore, but we have not investigated, how this conclusion depends on the particle sampling scheme for the conversion from the hydrodynamics to microscopic transport. How the application of an unfolding procedure affects the results and an analysis of the decorrelation over the different stages of heavy ion reactions are detailed questions that we did not attempt to answer in this study, since there are important missing pieces in the modeling approaches. 

As long as it is not properly understood how the hadronization of the fireball distorts the event-by-event correlation between initial state geometry and final state observables, the idea of event shape engineering cannot be used for precision studies of geometric effects on other probes, such as jets. To gain more insight into the event-by-event correlation between initial state geometry and measured azimuthal momentum anisotropy, improved theoretical approaches are needed that can describe finite size effects on the momentum correlations throughout the whole fireball evolution. In addition, a better microscopic understanding of the transition from locally equilibrated matter to hadronic matter out of equilibrium would be desirable.

\section*{Acknowledgements}
\label{ack} We gratefully acknowledge the Open Science Grid for making computing resources available. We thank Dirk Rischke for providing the hydrodynamics code. H.P. acknowledges funding by the Helmholtz Association for the Young Investigator Group VH-NG-822. This work was supported in part by U.S. Department of Energy grant DE-FG02-05ER41367.

\end{document}